\documentclass[aps,pra,groupedaddress,amssymb,amsmath,showpacs,twocolumn]{revtex4}
\usepackage{graphicx}

\begin{document}

\title{Bose-Einstein condensation for trapped  atomic polaritons in a  biconical waveguide cavity}
\author{I. Yu. Chestnov}
\author{A. P. Alodjants}
\email[Electronic address]{:alodjants@vlsu.ru}
\author{S. M. Arakelian}
\affiliation{Department of Physics and Applied Mathematics, Vladimir State University named after
A. G. and N. G. Stoletovs, Gorky str. 87, 600000, Vladimir, Russia}
\author{J. Klaers, F. Vewinger, M. Weitz}
\affiliation{Institut fur Angewandte Physik der Universit\"at Bonn, Wegelerstra\ss e 8, 53115 Bonn,
Germany}

\begin{abstract}
We study the problem of high temperature Bose-Einstein condensation (BEC) of atom-light polaritons
in a waveguide cavity appearing due to interaction of two-level atoms with (non-resonant) quantized
optical radiation, in the strong coupling regime, in the presence of optical collisions (OCs) with
buffer gas particles. Specifically, we propose  a special biconical waveguide cavity (BWC),
permitting localization and trapping of low branch (LB) polaritons imposed by the variation of the
waveguide radius in longitudinal direction. We have shown that   critical temperature of BEC
occurring in the system can be high enough -- few hundred Kelvins; it is connected with photon-like
character of LB polaritons and strongly depends on waveguide cavity parameters. In the case of a
linear trapping potential we obtain an Airy-shaped  polariton condensate wave function which, when
disturbed out of equilibrium, exhibits small amplitude oscillations with the characteristic period
in the picosecond domain.
\end{abstract}
\maketitle

\section{INTRODUCTION}

The investigation of quantum and statistical properties of Bose-gases in low and especially in one
dimension  (1D) has evoked indefatigable  interest in atomic optics and condensed matter physics
for the last few decades,   see e.g. \cite{1,2,3,4,5,6}. In particular, at finite temperatures for
a 1D or 2D ideal Bose gases a true Bose-Einstein condensation (BEC) can only be reached in the
presence of a suitable trapping potential \cite{4}, and the critical temperature for the phase
transition depends on the shape of trapping potential, which is usually harmonic in practice, cf.
\cite{3} and \cite{4}. Low-dimensional systems have been studied using atoms in highly deformed
traps \cite{10} where effects of dimensional reduction become important \cite{11}. Alternative
systems are bosonic quasiparticles, where light and matter are coupled in a coherent way -- see
e.g. \cite{12}. Here quantum and statistical properties of light Bose-quasiparticles, like excitons
\cite{13_}, magnons \cite{14_} and polaritons (see e.g. \cite{13,14,15,16}) have been considered.
For example, exciton-polaritons occurring  in quantum well structures placed in microcavities  can
be treated as a \textit{2D gas} of bosonic particles having an effective mass which is many orders
smaller than the mass of an electron in vacuum. This allows to study relatively high-temperature
phase transitions in low-dimensional bosonic systems. Recently, the evidence of a
Kosterlitz-Thouless phase transition -- cf. \cite{2}, superfluid behaviour  of exciton-polaritons
in such systems, has been reported by many labs, \cite{13,14}. However, in the current
semiconductor structures the thermalization time is in the picoseconds domain and comparable with
the particle lifetime, thus we deal here with non-equilibrium ``condensates'' \cite{15}, for which
dissipative and optical pumping effects play a crucial role, cf. \cite{16}. The characteristic
temperatures for these condensates are a few ten Kelvins, which is far above the atomic
condensates, but still far below room temperature.

Recently room-temperature Bose-Einstein-condensation of photons has been observed \cite{15a,15_},
where the photons are confined  in a 2D  curved-mirror optical microresonator filled with a dye
solution. Thermalization of  the photon gas is established by thermal contact between the photons
and the dye solution, using repeated absorption and re-emission processes in the dye solution
\cite{15b}. Frequent collisions, on a timescale much faster than the excited state lifetime lead to
a decoupling of photons and dye molecules, thus the relevant particles are not polaritons but
photons (see \cite{15_,15a}) in these experiments.

In the present paper we discuss a different approach to reach a high (room and beyond) temperature
phase transition with mixed matter-field states  - polaritons in an atomic medium. In particular,
dressed-sate polaritons, where a light field is coherently and strongly coupled to a two-level
atom, leading to a bosonic quasiparticle (in a suitable limit) are attractive candidate due to
their potentially longer lifetime, cf, \cite{17,18}. In this system thermal equilibrium of coupled
atom-light (dressed) states can be achieved experimentally within a nanosecond domain and is
limited  by the natural lifetime of the two-level atoms. Here, optical collisions (OCs) with buffer
gas atoms can lead to thermalization, experimentally evidence for a thermal quasi-equilibrium of
coupled atom-light states has been found \cite{19,20}. The obtained time of thermalization have
been about ten times shorter than the natural lifetime at full optical power. Here, the limited
available power in the light field prevented a full thermalization. To overcome this problem, in
\cite{21,22} special metallic waveguides of various configurations with the length up to a few
millimeters have been considered for trapping the polaritons inside, similar to waveguides and
closed resonators examined for the confinement of microwave irradiation, cf. \cite{23_}. The
lifetime of photon-like polaritons trapped in the waveguide can be longer than the thermalization
time, and is mainly determined by the cavity $Q$-factor. Thus we expect in such a waveguide that
for a large and negative atom-field detuning a high-temperature phase transition to a superradiant
state of the polaritons can be reached \cite{22}.

In the paper we study thermodynamic and critical properties of dressed state polaritons trapped in
a biconical waveguide, i.e. metallic microtubes with different geometry \cite{23}. In this case one
can study the thermodynamics of a 1D quantum gas, which exhibits a high temperature phase
transition to a BEC in the case of a trapping potential which is more confining than harmonic
\cite{3}. For this reason we consider waveguides with a biconical shape, which leads to a trapping
potential in propagation direction. Throughout the paper, we assume the strong coupling limit, i.e.
the eigenstates of the coupled atom-light system are treated as polaritons. In Sec. II we describe
the waveguide cavity model for confinement of the optical mode. To be more specific, we consider
field properties in an empty lossless metallic biconical waveguide cavity (BWC) where the photons
are confined in transversal and trapped in longitudinal dimension. In Sec. III we give a quantum
description of the atom-field interaction in the waveguide, and the problem of specifying waveguide
parameters in the limit of weak trapping is considered. Thermodynamic properties  of the 1D ideal
gas of photon-like polaritons are examined, where we find a phase transition to a BEC state at
experimentally feasible temperatures. In Sec. IV we discuss the dynamical properties of condensed
LB polartions. In the conclusion, we summarize the results obtained.

\section{BICONICAL WAVEGUIDE   CAVITY}

\begin{figure}
\includegraphics[scale=0.28]{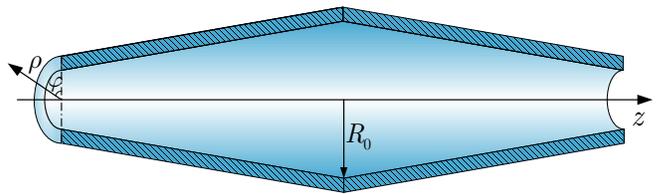}
\caption{\label{fig1}(Color online) Biconical waveguide for photon (and polariton) trapping. Shown
is a waveguide with a radius variation following the form $R(z) = R_0 \left( 1 - \alpha \left| z
\right| \right) $, which is obtained from eq.\eqref{eq:diameter} for $\nu = 1$ in the limit
of a small value of $\alpha$. In this case the radius variation within the limited waveguide length
is roughly linear. The radius at the waveguide center $z=0$ is taken to be close to $\lambda/2.61$,
yielding a low frequency cutoff close to the atomic resonance.}
\end{figure}

We start by describing the problem of photon trapping in the empty BWC, sketched in
Fig.\ref{fig1}. We assume metallic boundary conditions, and for the case of simplicity the cavity
is assumed to be lossless. The radius of the cylindrically symmetric waveguide depends on $z$
following
\begin{equation} \label{eq1_}
R(z)=\frac{R_{0} }{F\left( \left| z \right|  \right)  }
\end{equation}
with $F\left( 0 \right) =1$ and $F \left( \left| z \right| \right) $ being a function increasing
monotonically with $\left| z \right|$, i.e. $\frac{\partial F}{\partial z} > 0 $  for $z >
0$, similar to bottle resonators based on glass materials (and with relatively large diameter), see
Ref. \cite{21}. We here focus on waveguides with a maximal diameter that is close to the wavelength
of the optical field. In this limit it is possible to guarantee a spacing of the transverse modes
above the thermal energy, which allows for the transverse mode quantum numbers to be frozen. The
system becomes effectively one-dimensional, with a continuum of longitudinal modes above a low
frequency cutoff. For a suitable density of states BEC then becomes possible in this one
dimensional situation. For a related, though two-dimensional, situation see \cite{15}. We consider
biconical waveguides, because fabrication of a conical hole can be done by e.g.\ ion beam etching
or laser drilling, and then combining two of those holes to get the required cavity. Alternatively,
self-assembling nanotubes can be used, which have been shown to be suitable for implementation in
the experiment \cite{21}.

We use the adiabatic approximation, assuming that $R(z)$ is slowly varying with the $z$ coordinate,
i.e. that the condition
\begin{equation} \label{eq2_}
\left| \frac{dR(z)}{dz} \right| \ll 1
\end{equation}
is fulfilled. In other words, we suppose that the angle between waveguide and the $z$-axis is small
throughout the length of the waveguide -- see Fig.\ref{fig1}.

The field properties in such a system can be described by the wave (scalar Helmholtz) equation for
the vector-potential $\Phi (x,y,z)$ (we consider \textit{TM-modes} of polarization only) -- cf.
\cite{25}:
\begin{equation} \label{eq3_}
\Delta \Phi (x,y,z)+k^{2} \Phi (x,y,z)=0,
\end{equation}
where we assume that the field evolves in time as $e^{i\omega t} $, $k=\omega /c$ is the wave
vector. We can interpret $\Phi (x,y,z)\equiv \Phi _{l} (x,y,z)$ as a photonic wave function in the
waveguide, where the index $l$ numerates a set of quantum numbers corresponding to the solution of
Eq.\eqref{eq3_} in terms of special functions.

Both electric ($E_{\rho } ,E_{\varphi } $ and $E_{z} $) and magnetic ($H_{\rho } ,H_{\varphi } ,H_{z} $)
field components can be found by using standard expressions for the $\Phi (x,y,z)$-function
derivatives in cylindrical coordinates -- see e.g. \cite{25}. For TM-modes the $z$ component of the
magnetic field is $H_{z} =0$, and the metallic boundary condition imposes that the electric field
$E$ has only a normal component on the waveguide surface.

Taking into account axial symmetry of the waveguide in Fig.\ref{fig1} it is efficient to rewrite
Eq.\eqref{eq3_} in cylindrical coordinates $z$, $\rho $ and $\varphi $ as:
\begin{equation} \label{eq4_}
\frac{1}{\rho } \frac{\partial }{\partial \rho } \left(\rho \frac{\partial \Phi }{\partial \rho }
\right)+\frac{\partial ^{2} \Phi }{\partial z^{2} } +\frac{1}{\rho ^{2} } \frac{\partial ^{2} \Phi
}{\partial \varphi ^{2} } +k^{2} \Phi =0,
\end{equation}
where the wave vector $k$ is defined as:
\begin{equation} \label{eq5_}
k^{2} =k_{\bot }^{2} +k_{z}^{2} .
\end{equation}

We are looking for the solution of Eq.\eqref{eq4_} in the form:
\begin{equation} \label{eq6_}
\Phi (\rho , z , \varphi)=\Psi (z)\Theta (\rho ,z)e^{im\varphi },
\end{equation}
where $m$ is an integer (azimuthal quantum) number, and the functions $\Theta (\rho ,z)$ and  $\Psi
(z)$ comply with the equations
\begin{equation} \label{eq7_}
\frac{\partial ^{2} \Theta }{\partial \rho ^{2} } +\frac{1}{\rho } \frac{\partial \Theta }{\partial
\rho } +(k_{\bot }^{2} (z)-\frac{m^{2} }{\rho ^{2} } )\Theta =0,
\end{equation}
\begin{equation} \label{eq8_}
\frac{\partial ^{2} \Psi (z)}{\partial z^{2} } +k_z^{2} \Psi (z)=0.
\end{equation}
In Eq.\eqref{eq7_} the terms containing derivatives on longitudinal coordinate $z$ are omitted due
to the adiabaticity condition \eqref{eq2_}. The solution for the radial distribution $\Theta (\rho
,z)$ can be expressed in terms of Bessel functions,
\begin{equation} \label{eq9_}
\Theta (\rho ,z)=J_{m} (k_{\bot } \rho ),
\end{equation}
where the wave vector component $k_{\bot } $ is quantized in this case. Omitting lengthy but
straightforward calculations  for the $k_{\bot } $ component of the wave vector for TM-modes in the
waveguide, we obtain:
\begin{equation} \label{eq10_}
k_{\bot ,{mp}} (z)=\frac{g_{mp} }{R(z)} =k_{\bot ,{mp}}^{(0)} F\left( \left| z \right|
\right) ,
\end{equation}
where $k_{\bot ,{mp}}^{(0)} =\frac{g_{mp} }{R_{0} } $ is a transversal wave vector
component; $g_{mp} $ is $p$-th zero of the Bessel function $J_{m} (x)$ of the $m$-th order
\cite{27}. In \eqref{eq10_} we have used the metallic boundary conditions for electric field
components, $J_{m} (k_{\bot } R(z))=0$. For $p=1$ and $m=0$ we find the cavity transverse ground
state mode, yielding $k_{\perp ,01}(z)=\frac{g_{01}}{R(z)}$ with $g_{01} \approx 2.405$.

Equation \eqref{eq5_} implies a dispersion relation of the form $\omega _{\rm ph} = ck= c
\sqrt{\left( k^{(0)}_{\perp {\rm ,mp}} F(\left|z \right| )\right)^{2} +k^{2}_{z}}. $ In the
approximation $k_{\perp, {mp}}(z) \gg k_{z}$ we arrive at
\begin{equation}\label{eq11}
\omega _{\rm ph}  \simeq  c k^{(0)}_{\perp {\rm ,mp}} F(\left|z \right| ) + \frac{c k^{2}_{z}}{2
k^{(0)}_{\perp {\rm ,mp}} F(\left|z \right| ) }.
\end{equation}
In this limit the second term is much smaller than the first term and we assume that in the
smaller, kinetic energy term we can set $F( \left| z \right| )=1$. This approximation is valid in
the limit of a not too large overall diameter variation of the waveguide. We arrive at a photon
energy
\begin{equation}\label{eq12}
E_{\rm ph}=m_{\rm ph}c^2 + \frac{ \left( \hbar k_z \right)^2}{2 m_{\rm ph}} + V_{\rm ph}(z),
\end{equation}
where we have defined an effective photon mass $m_{\rm ph}=\frac{\hbar k^{(0)}_{\perp ,mp}}{c}$ (with
$\omega_{\rm cutoff}=m_{\rm ph}c^2/ \hbar$ as the low frequency cutoff) and an effective photon
trapping potential $V_{\rm ph}(z)=\hbar ck^{(0)}_{\perp} \left( F( \left| z \right| ) -1 \right)$ in
analogy to \cite{15_}. The system becomes formally equivalent to a one-dimensional ideal gas of
massive particles moving along the $z$-axis under confinement in the potential $V_{\rm ph}(z)$.

In the following we assume a dependence
\begin{equation}\label{eq:diameter}
 F( \left| z \right| )=1 + \alpha \left| z \right| ^{\nu},
\end{equation} corresponding to a waveguide diameter variation along the $z$-axis of
$R(z) = R_0 / \left( 1 + \alpha \left| z \right| ^{\nu} \right) $, with $\nu \geqslant 0$ and
$\alpha \geqslant 0$, we arrive at a potential of the form $V_{\rm ph}(z)=\hbar c k^{(0)}_{\perp} \alpha
\left| z \right| ^{\nu}= m_{\rm ph}c^2 \alpha \left| z \right| ^{\nu}$. True BEC is expected to be
possible in the 1D system, if the potential is more confining than parabolic, i.e. $\nu < 2$
\cite{3}. We here mostly are interested in the case of $\nu =1$, for which $R(z) = R_0 / \left( 1 +
\alpha \left| z \right| \right) $, which yields a linear trapping potential following $V_{\rm
ph}(z) = m_{\rm ph}c^2 \alpha \left| z \right|$, see Fig. \ref{fig1}. For the case of a relatively
small variation of the waveguide diameter over the length $l$ of the waveguide, i.e. $\frac{1}{2}
\alpha l \ll 1$, one may use a linear variation of the diameter $R(z) \approx R_0 \left( 1 - \alpha
\left| z \right| \right) $, which gives the desired potential in first order approximation
\cite{21}. Such a purely biconical design can be easier to fabricate experimentally.

Given the dispersion relation \eqref{eq12} the longitudinal wavefunction  $\Psi (z)$,
Eq.\eqref{eq8_}, follows the Schr\"odinger equation
\begin{equation} \label{eq14_}
\frac{\partial ^{2} \Psi (z)}{\partial z^{2} } +\frac{2m_{\rm ph}}{\hbar^2} \left(E- U_{\rm ph}|z|^{\nu } \right)\Psi (z)=0,
\end{equation}
where $E=E_{\rm ph}-m_{\rm ph}c^2$  is the shifted  photon energy in the cavity, and  we made
definition $U_{\rm ph}=m_{\rm ph}c^2 \alpha$.

The solutions $\Psi _{n}(z)$ of the Schr\"odinger equation \eqref{eq14_} can be given in terms of
special functions, see e.g. \cite{27}. In the quasi-classical approach the energy spectrum of photonic field in the waveguide can be
obtained from familiar Bohr-Sommerfeld quantization principle and reads (see e.g. \cite{30})
\begin{equation} \label{eq15}
E_{n} =\hbar \omega _{\nu } (n+1/2)^{\frac{2\nu }{\nu +2} } ,
\end{equation}
with the characteristic frequency of photonic ``particle'' oscillation, i.e. trapping frequency in
the cavity
\begin{equation} \label{eq16_}
\omega _{\nu } =\left(\frac{\pi U_{\rm ph}^{1/\nu } }{2^{3/2} m_{\rm ph}^{1/2} \hbar ^{(2-\nu
)/2\nu } I(\nu )} \right)^{\frac{2\nu }{\nu +2} } .
\end{equation}
Here we have defined $I(\nu)=\int_{0}^{1}\left(1-|t|^{\nu } \right)^{1/2} dt $.

A characteristic length for the waveguide cavity can be obtained by looking at the turning points
$z=\pm z_{c} $ for the photonic ground state mode, given by
\begin{equation} \label{eq16}
z_{c} =\left({E\mathord{\left/ {\vphantom {E U_{ph} }} \right. \kern-\nulldelimiterspace} U_{ph} }
\right)^{1/\nu } =\left(\frac{\pi }{4I(\nu )} \right)^{\frac{2 }{2+\nu } } d_{\nu },
\end{equation}
where $d_{\nu } =\left(\frac{\hbar^2 }{2m_{ph}^{2} c^2\alpha } \right)^{{1\mathord{\left/
{\vphantom {1 (\nu +2)}} \right. \kern-\nulldelimiterspace} (\nu +2)} } $ defines a characteristic
longitudinal  scale of  localization of the ground state ($n=0$). From this we can define the
condition for the validity of the quasiclassical approach, given as (cf. \cite{26})
\begin{equation} \label{eq17}
k_{z} d_{\nu }\gg {(\nu \mathord{\left/ {\vphantom {(\nu  2}} \right. \kern-\nulldelimiterspace} 2}
)^{1/3} .
\end{equation}
Relation \eqref{eq17} can be satisfied for a given $k_z$ by adjusting the waveguide parameter
$\alpha $, cf. \eqref{eq2_}.

\section{THERMODYNAMICS OF POLARITONS IN THE WAVEGUIDE CAVITY}\label{section_numerics}

The description of atomic polaritons in a small volume cavity can be given in the similar way as
represented in \cite{18,22}, i.e. by using Holstein-Primakoff transformation for atomic
excitations. The Hamiltonian of the total system of atom and quantized field is given by $H=H_{\rm rad}+H_{\rm at} +H_{\rm int} $, where $H_{\rm rad} $ characterizes non-interacting photons in the waveguide,
$H_{\rm at} $ is a Hamiltonian of atomic ensemble, and $H_{int} $ is responsible for the
interaction of $N_{\rm at} $ two-level atoms with quantized optical field in the cavity.  In
momentum representation the Hamiltonian $H$ may be written as
\begin{multline} \label{eq20_}
H=\hbar \sum _{\vec{k}} \left(\omega _{\rm ph} \hat{f}_{\vec{k}}^{\dag } \hat{f}_{\vec{k}} +\omega _{\rm at} \hat{\phi }_{\vec{k}}^{\dag } \hat{\phi }_{\vec{k}} +\kappa \left(\hat{f}_{\vec{k}}^{\dag } \hat{\phi }_{\vec{k}} +\hat{\phi }_{\vec{k}}^{\dag } \hat{f}_{\vec{k}} \right)\right) - \\
- \frac{\hbar \kappa }{2N_{\rm at} } \sum _{kk'q}\left(\hat{f}_{k+q}^{\dag } \hat{\phi
}_{k'-q}^{\dag } \hat{\phi }_{k} \hat{\phi }_{k'} +\hat{\phi }_{k}^{\dag } \hat{\phi }_{k'}^{\dag }
\hat{\phi }_{k'-q} \hat{f}_{k+q} \right),
\end{multline}
where $\hat{f}_{\vec{k}} $ ($\hat{f}_{\vec{k}}^{\dag } $) is the annihilation (creation) operator
for the photons absorbed (or emitted), $\hat{\phi }_{\vec{k}} $ ($\hat{\phi }_{\vec{k}}^{\dag } $)
is the annihilation (creation) operator that characterizes  excitations (polarization) of a
two-level atomic ensemble and obeys usual commutation relation for a Bose system, $\kappa
=\left(\frac{|\wp _{ab} |^{2} \omega _{L} N_{\rm at} }{2\hbar \varepsilon _{0} V_{M} }
\right)^{1/2} $ is the collective  atom-field interaction strength, $\wp _{ab} $ is atomic dipole
matrix element, $V_{M} $\textit{ }is an effective volume of mode occupation within the region of
atom-field interaction that can be defined as $V_{M} =\int _{\rm cavity} \frac{\left|\Phi
(r)\right|^{2} }{\max \left(\left|\Phi (r)\right|^{2} \right)} d^{3} r$, where  $\max
\left(\left|\Phi (r)\right|^{2} \right)$ is maximal value of the square of the wave function, cf.
\cite{22,24}. Here we have assumed that the dipole matrix element $\wp
_{ab}$ is independent of the mode. This simplification is justified, as the exact transition rates
do not influence the thermodynamic properties of the system. The nonlinear part of the Hamiltonian
$H$, i.e. the last term in \eqref{eq20_} characterizes two-body polariton interaction processes due
to atomic saturation effects. The polaritonic dispersion is determined by the photon dispersion
$\omega _{\rm ph} (k)$ and the atomic excitation dispersion, which is described by
\begin{equation} \label{eq22_}
\omega _{\rm at} \equiv \omega _{\rm at} (k)=\omega _{0} +\frac{\hbar k_{z}^{2} }{2m_{\rm at} } ,
\end{equation}
where $\omega _{0} $ is the atomic transition frequency.

If a quantum field intensity is not too high (the number of photons is essentially smaller than the
number of atoms), we can assume that the corresponding dispersion relation for polariton states is
not modified compared to the uncoupled case. Thus, we can  use a polariton basis to diagonalize the
total Hamiltonian in \eqref{eq20_} by using the unitary transformation
\begin{subequations} \label{eq23_}
\begin{eqnarray}
\hat{\Xi }_{1,\vec{k}} =X_{k} \hat{f}_{\vec{k}} +C_{k} \hat{\phi }_{k}, \\
\hat{\Xi }_{2,\vec{k}} =X_{k} \hat{\phi }_{k} -C_{k} \hat{f}_{\vec{k}},
\end{eqnarray}
\end{subequations}
where the introduced annihilation operators $\hat{\Xi }_{1,\vec{k}} $, $\hat{\Xi }_{2,\vec{k}} $
characterize polaritons in the atomic medium, corresponding to two types of elementary excitations
which in low density limit satisfy usual boson commutation relations, cf.\cite{22}. Parameters
$X_{k} $ and $C_{k} $ are real Hopfield coefficients satisfying condition $X_{k}^{2} +C_{k}^{2}
=1$, which determines the contribution of the photon ($C_{k} $) and atomic excitation ($X_{k} $)
fraction to the polariton annihilation  operators \eqref{eq23_} according to
\begin{subequations} \label{eq24_}
\begin{eqnarray}
X_{k} =\frac{1}{\sqrt{2} } \left(1+\frac{\delta _{k} }{\sqrt{4\kappa ^{2} +\delta _{k}^{2} } } \right)^{1/2}, \\
C_{k} =\frac{1}{\sqrt{2} } \left(1-\frac{\delta _{k} }{\sqrt{4\kappa ^{2} +\delta _{k}^{2} } }
\right)^{1/2},
\end{eqnarray}
\end{subequations}
where $\delta _{k} =\omega _{\rm ph} -\omega _{\rm at}\approx \Delta +\frac{\hbar k_{z}^{2}
}{2m_{\rm ph} } +\frac{V_{\rm ph}^{} (z)}{\hbar }$ is the frequency mismatch. Here
$\Delta=\omega_\text{L}-\omega_0$ is the atom-field detuning, where $\omega_{L}$ is laser light frequency which is taken to be close to $\omega_{\rm cutoff}$. In the presence of photon
trapping $X_{k} $ and $C_{k} $ parameters depend on waveguide longitudinal coordinate $z$.

We confine our analysis to polaritons of the lower branch (LB polaritons), and we ignore the
effects of interactions between lower and upper polariton branches. Taking into account
quasiclassical approach (cf. \eqref{eq17}) for the photonic field in the BWC
we find the general conditions for the observation of a BEC of (lower branch) polaritons,
\begin{equation} \label{eq25_}
\Delta E_{n} \ll k_{B} T\ll\hbar \Omega _{R0},
\end{equation}
where $\Omega _{R0} =\left(\Delta ^{2} +4\kappa ^{2} \right)^{1/2} $ is \textit{zero-momentum} Rabi
splitting frequency between the lower and upper polariton branch. The first constraint in
\eqref{eq25_} represents the condition for a quasiclassical limit where the energy spacing $\Delta
E_{n} \sim \hbar \omega _{\nu}$ (see \eqref{eq15}) of the quantized photonic states is essentially
smaller than the thermal energy and the states may be treated as a continuum, cf. \cite{3}, where
the energy levels are populated according to a Bose-Einstein distribution in thermal equilibrium.
The second condition implies that the thermal energy is not enough to excite lower branch
polaritons to the upper branch, essentially allowing us to neglect the upper branch.

Taking Eqs. \eqref{eq23_}, \eqref{eq24_} into account from Eq.\eqref{eq20_} we can obtain LB
Hamiltonian $H_{\rm LB} $ in the form
\begin{equation} \label{eq26_}
H_{\rm LB} =\hbar \sum _{\vec{k}} \Omega _{\vec{k}} \hat{\Xi }_{2,\vec{k}}^{\dag } \hat{\Xi
}_{2,\vec{k}} +\sum _{kk'q}U_{\vec{k}\vec{k}'\vec{q}} \hat{\Xi }_{2,k+q}^{\dag } \hat{\Xi
}_{2,k'-q}^{\dag } \hat{\Xi }_{2,k} \hat{\Xi }_{2,k'}  ,
\end{equation}
where $\Omega _{\vec{k}} =\frac{1}{2} \left[\omega _{\rm at} +\omega _{\rm ph} -\Omega _{R}
\right]$ determines the dispersion relation for LB polaritons; $\Omega _{R} =\left(\delta ^{2}
+4\kappa ^{2} \right)^{1/2} $ is the Rabi splitting frequency that determines the  gap between
upper and lower states.  The gap is minimal and equal to $\Omega _{R0} $ taken for $k_{z} =0$ at
the center of the trap (waveguide) at $z=0$. In Eq.\eqref{eq26_} we give the definition
$U_{\vec{k}\vec{k}'\vec{q}} =\frac{\hbar \kappa }{2N_{\rm at} }
\left(C_{\left|\vec{k}+\vec{q}\right|} X_{\vec{k}'} +C_{\vec{k}'} X_{\left|\vec{k}+\vec{q}\right|}
\right)X_{\left|\vec{k}'-\vec{q}\right|} X_{\vec{k}} $ that determines two-body polariton-polariton
scattering processes. The effective mass of the LB polaritons is found to be
\begin{widetext}
\begin{equation} \label{eq27_}
m_{\rm pol} \equiv \hbar \left(\left. \frac{\partial ^{2} \Omega _{\vec{k}} }{\partial k_{z}^{2} }
\right|_{k_{z} =0} \right)^{-1} =\frac{2m_{\rm at} m_{\rm ph} \Omega _{Rz} (z)}{\left(m_{\rm at}
+m_{\rm ph} \right)\Omega _{Rz} (z)-\left(m_{\rm at} -m_{\rm ph} \right)\left(\Delta +{V_{\rm
ph}^{} (z)\mathord{\left/ {\vphantom {V_{\rm ph}^{} (z) \hbar }} \right. \kern-\nulldelimiterspace}
\hbar } \right)}  ,
\end{equation}
\end{widetext}
where $\Omega _{Rz} (z)=\left(\left(\Delta +{V_{\rm ph}^{} (z)\mathord{\left/ {\vphantom {V_{\rm
ph}^{} (z) \hbar }} \right. \kern-\nulldelimiterspace} \hbar } \right)^{2} +4\kappa ^{2}
\right)^{1/2} $ is the $z$ dependent Rabi splitting frequency.

To be more specific we examine the interaction between a quantized field and rubidium atoms, which
are treated as a two-level system. The transition frequency is taken as the weighted mean of the
rubidium \textit{D-}lines, $\omega _{0} /2\pi \simeq 382$ THz, and the laser is taken to be red
detuned by $\left|  \Delta \right| /2\pi \geqslant 11$ THz , $\Delta < 0$ \cite{19,20}. In the experiments reported the thermal
energy ($k_{B} T$) for the atoms at ambient temperatures ($T=530$ K) is about the frequency of Rabi
splitting $\Omega _{R0} /2\pi \simeq 11$ THz in energy units, thus condition \eqref{eq25_} has not
been completely achieved yet \cite{19,20,21,22}. We consider the perturbative limit when $\Omega
_{0} ,{\rm \; }\kappa \ll\left|\Delta \right|$, that is $\Omega _{R0}/2\pi \approx \left|\Delta
\right|/2\pi \geqslant 11$ THz.  For negative detuning ($\Delta <0$) the LB polaritons with
relatively small momentum $k_{z}^{} $ at the center of the trap are photon-like with the mass
$m_{\rm pol} \approx m_{\rm ph} \frac{2\Omega_{R0}}{\Omega_{R0}+\left| \Delta \right|} \simeq 2.8
\times 10^{-36}$ kg, that is $\delta _{k} \simeq \Delta $, $X_{k} \approx 0,{\rm \; }C_{k} \approx
1,$ and $\hat{\Xi }_{2,\vec{k}} \simeq -\hat{f}_{\vec{k}} $, cf. \cite{22}.

The role of polariton-polariton scattering processes in Eq.\eqref{eq26_} is negligibly small in
this limit. At the bottom of the dispersion curve the polariton scattering parameter $U_{0} \equiv
\left. U_{\vec{k}\vec{k}'\vec{q}} \right|_{\vec{k}\vec{k}'\vec{q}=0} $ behaves as  $U_{0} \simeq
\frac{ \hbar \kappa ^{4} }{N_{\rm at} \left|\Delta \right|^{3} } $.  Hence in the limit of a large
atom-light detuning $\left|\Delta \right|$ (ratio $\kappa / \left| \Delta \right|$  is about
$0.057$, cf. \cite{20,22}) and for a macroscopically large number of atoms ($N_{\rm at} \gg 1$) the
gas of LB  polaritons can be treated as an ideal one.

\begin{figure}
\includegraphics[scale=0.5]{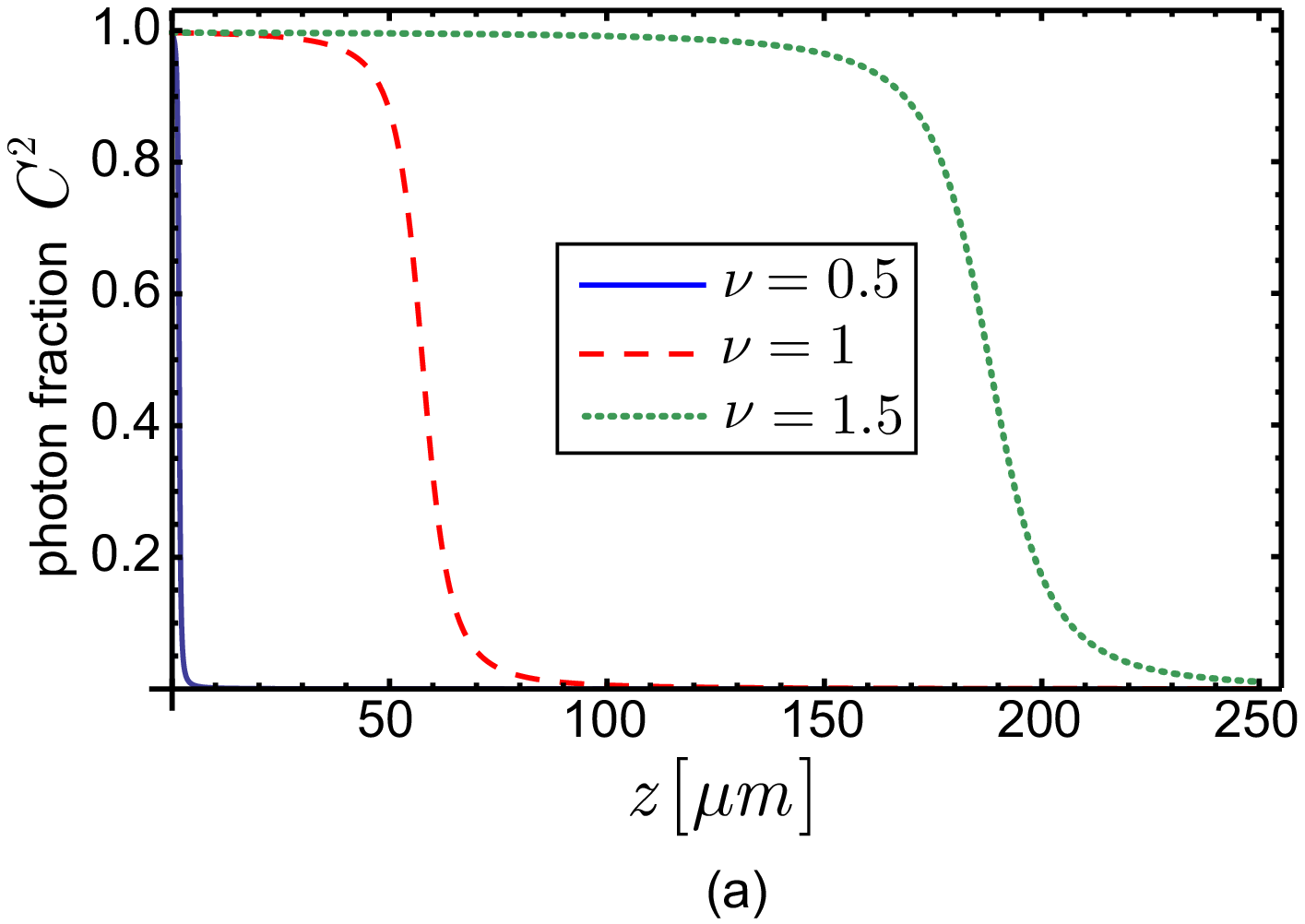}\ \
\includegraphics[scale=0.5]{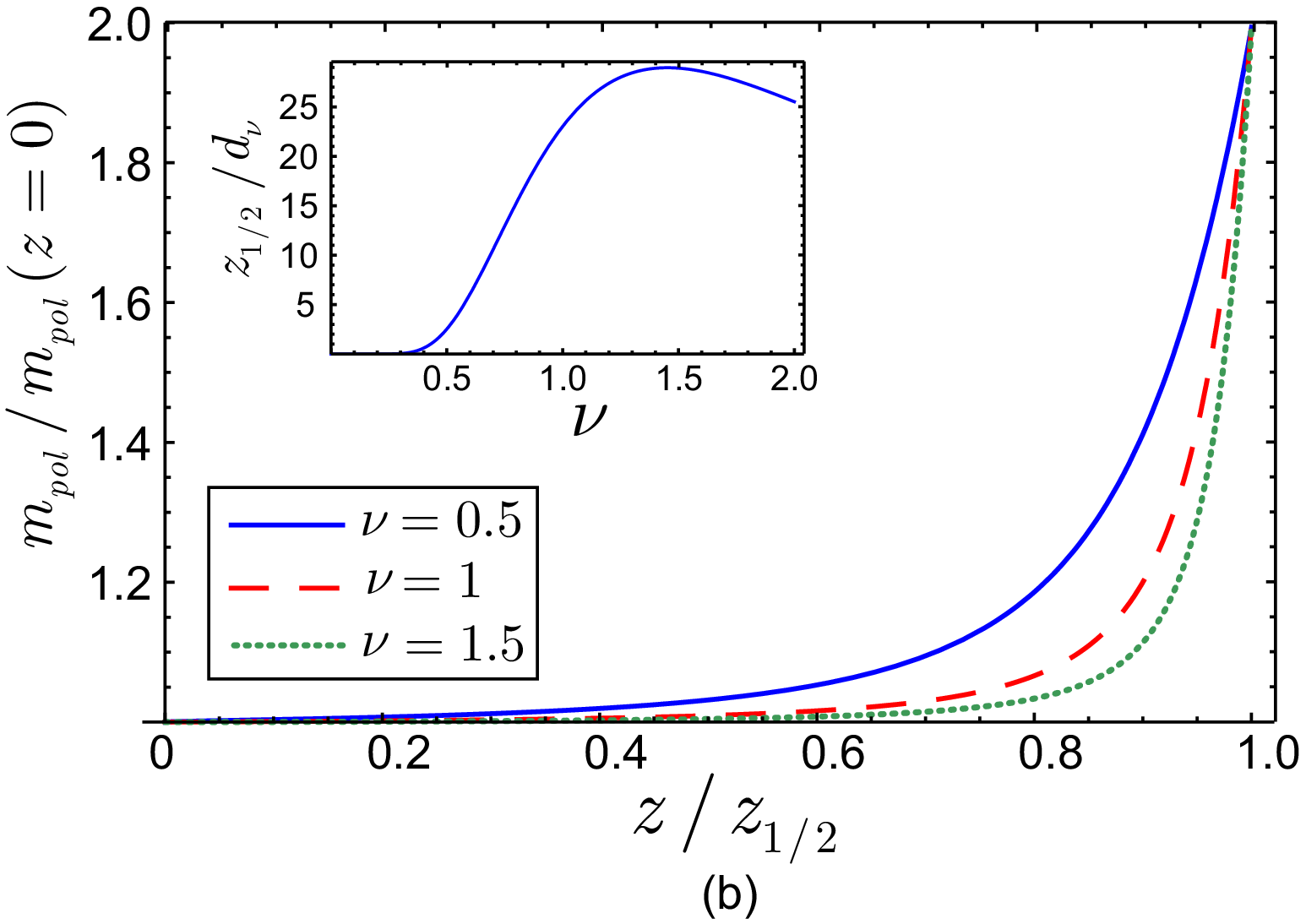}
\caption{\label{fig3}(Color online). (a) -- The polariton photon fraction (coefficient $C_{0}^{2}
$) and, (b) -- normalized  LB polariton mass $m_{\rm pol} $ as a function of $z$ for trapping power
parameter $\nu =0.5,{\rm \; \; 1,\; \; 1.5}$. The inset shows dependence of $z_{1/2}^{} /d_{\nu } $
versus parameter $\nu $. The parameters are $\Delta /2\pi =-11$ THz, $R_{0}\simeq \lambda/2.61 \approx 0.3$ $\mu{\rm m}$, $\alpha^{1/\nu} =0.0005$ $\mu {\rm m}^{-1} $. }
\end{figure}

The photonic fraction of the LB polaritons -- Hopfield coefficient $C_{0}^{2} \equiv C_{k_{z}
=0}^{2} $ -- and the mass of LB polaritons $m_{\rm pol} $ taken in the limit of zero momentum
$k_{z} =0$ as a function of $z$ are presented in Fig.\ref{fig3}a,b. With increasing coordinate $z$,
LB polaritons become more atom-like, and present half matter - half photon quasi-particles ($X_{0}
(z)=C_{0} (z)=2^{-1/2} $) with the mass $m_{\rm pol} \approx 2m_{\rm ph} $ at the distance $z_{1/2}
=\left({\hbar \left|\Delta \right|\mathord{\left/ {\vphantom {2\hbar \left|\Delta \right| U_{\rm
ph}}} \right. \kern-\nulldelimiterspace} U_{\rm ph} } \right)^{1/\nu } $, where photonic and atomic
dispersion lines cross (see Fig.\ref{fig5}a). From definition \eqref{eq27_} it follows that the
mass of polaritons becomes $z$ independent if condition  $V_{\rm ph}\left( z \right) \ll \hbar
\left| \Delta \right|$   is satisfied. Since $V_{\rm ph}\left( d_{\nu}\right) \sim \hbar \omega
_{\nu}$, the inequalities \eqref{eq25_}, already involve the above condition. In this limit an
effective width of the photonic mode localization $d_{\nu } $ is essentially smaller than the
characteristic length $z_{1/2} $. The ratio $z_{1/2}^{} /d_{\nu } $ for different trapping power is
outlined in the inset in Fig.\ref{fig3}b. For $\nu=1$ the region of photon mode localization is
more than ten times shorter than  characteristic length $z_{1/2} $, allowing to treat the LB
polaritons as nearly photonic. The variation of the polariton mass, see Fig.\ref{fig3}b, is small
for $z\ll z_{1/2} $, i.e. under condition \eqref{eq25_}, allowing to treat the LB polaritons as
particles with constant mass.

\begin{figure}
\includegraphics[scale=0.5]{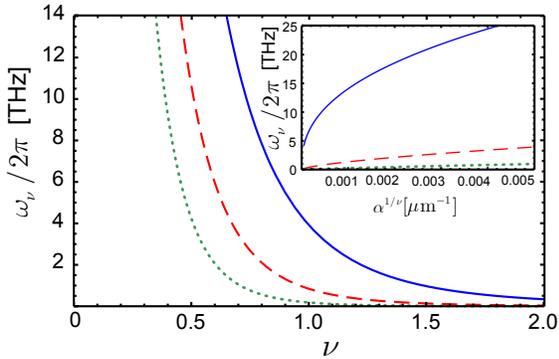}
\caption{\label{fig4}(Color online). Dependences of photonic trapping frequency $\omega
_{\nu}/2\pi$ versus trapping parameter $\nu $. Blue solid line corresponds to $\alpha^{1/\nu}
=0.005$ $\mu {\rm m}^{-1} $, red dashed -- to $\alpha^{1/\nu}  =0.0005$ $\mu {\rm m}^{-1} $, green
dotted -- to $\alpha^{1/\nu}  =0.00005$ $\mu{\rm m}^{-1} $. In the inset the dependence of $\omega
_{\nu}$ versus parameter $\alpha^{1/\nu}  $ for $\nu =0.5$ (blue solid line), $\nu =1$ (red dashed
line) and $\nu =1.5$ (green dotted line) is depicted.}
\end{figure}

\begin{figure}
\includegraphics[scale=0.45]{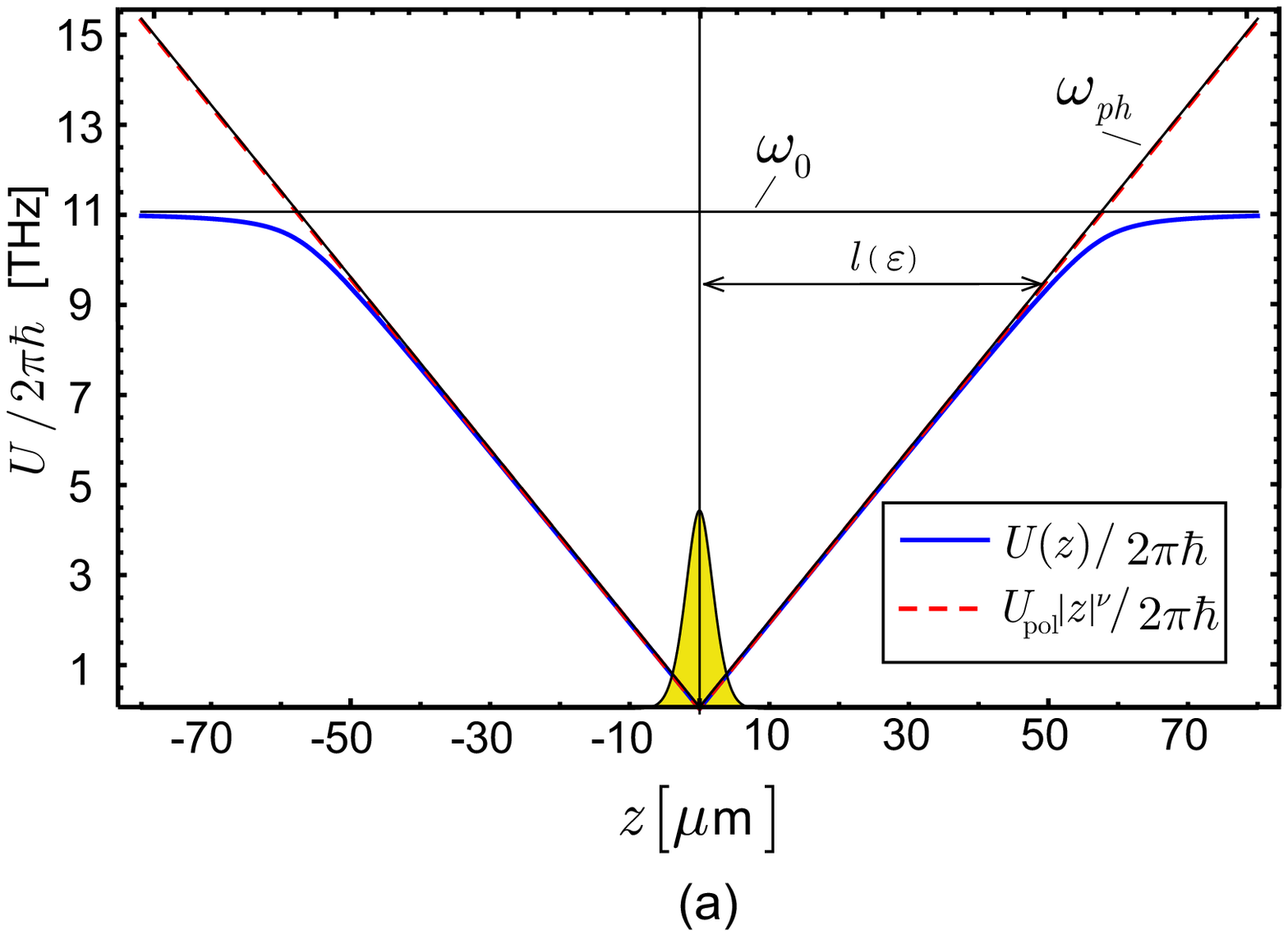}
\includegraphics[scale=0.6]{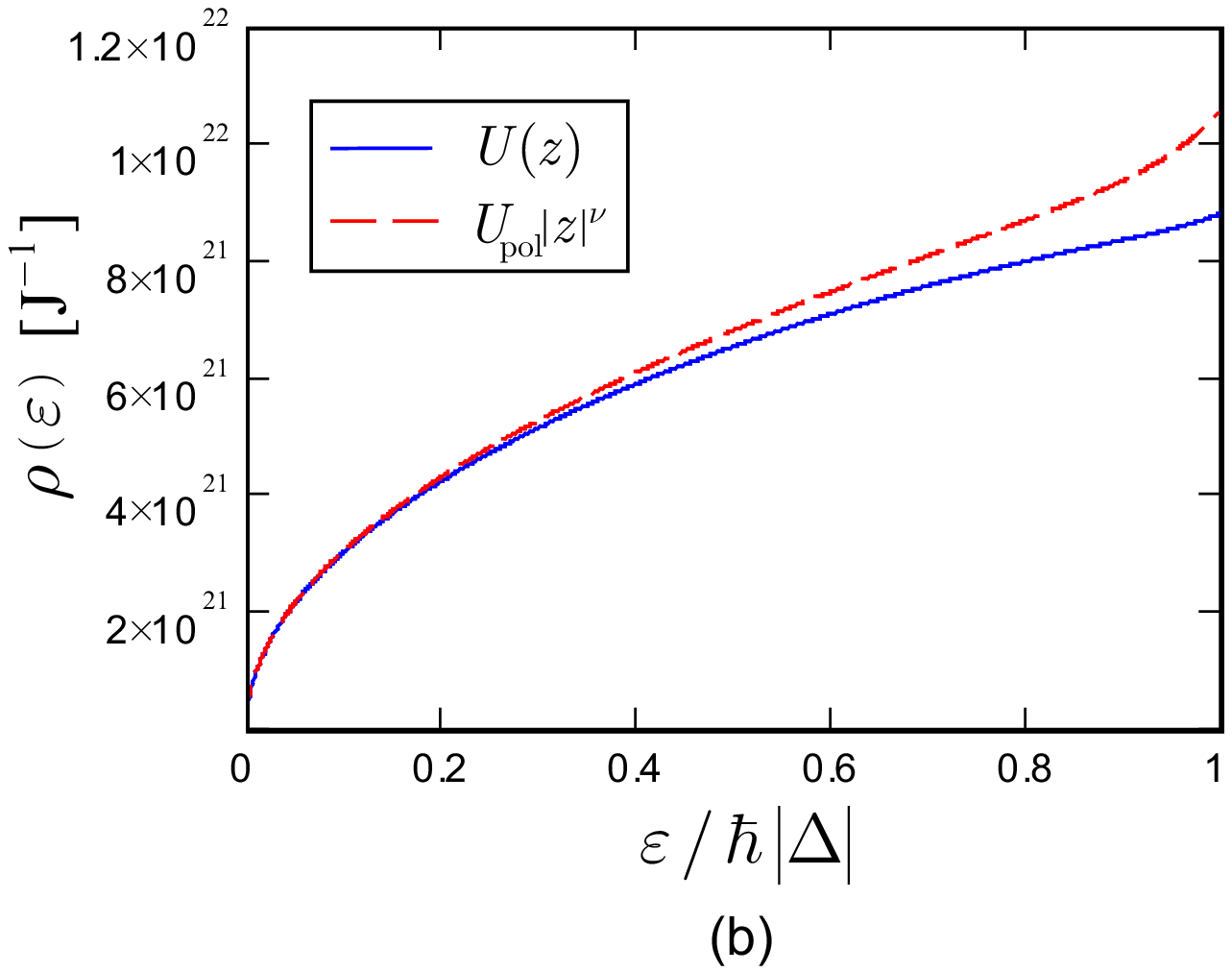}
\caption{\label{fig5}(Color online) (a) shows the energy of polariton trapping versus coordinate
$z$ taken for $\nu =1$; (b) shows the dependence of  density of states $\rho \left(\varepsilon
\right)$ on normalized polariton energy $\varepsilon /\hbar \left|\Delta \right|$. The solid curves
correspond to exact trapping potential \eqref{eq30_}, and the dashed ones are relevant to power-law
potential \eqref{eq31_}. The black solid lines correspond to atomic $\omega _{\rm at} $ and
photonic $\omega _{\rm ph} $ frequencies. The shaded (yellow) area corresponds to  LB polariton
ground state wave function -- see Sec. IV.}
\end{figure}

Figure \ref{fig4} shows the characteristic trapping frequency $\omega _{\nu}$ as a function of
parameter $\nu $ for the cavity, where red (dashed) curve corresponds to waveguide $\alpha
$-parameter used in Fig.~\ref{fig3}. For $\nu \ge 1$ the condition \eqref{eq25_}
can be fulfilled for any reasonable value of $\alpha $-parameter by choosing the atom-field
detuning $\left| \Delta \right|$. On the other hand,  $\omega _{\nu}$ increases within the domain
of  $0<\nu <1$ and the fulfillment of condition \eqref{eq25_} mainly depends on the given value of
curvature of the waveguide radius, i.e. on  $\alpha $-parameter and on experimentally accessible
atom-field detuning $\left|\Delta \right|$.

The results for the Rubidium system allow us to further simplify the Hamiltonian for the LB
polaritons: At small momenta $\frac{\hbar k_{z}^{2} }{2m_{\rm pol} } \ll \Omega_{R 0} $ it
approaches
\begin{equation} \label{eq29_}
H_{\rm LB} \approx \sum _{\vec{k}} \left(\frac{\hbar^{2} k_{z}^{2} }{2m_{\rm pol} }
+U(z)\right)\hat{\Xi }_{\vec{k}}^{\dag } \hat{\Xi }_{\vec{k}} ,
\end{equation}
where $U(z)$ is an effective trapping potential for polaritons defined as
\begin{equation} \label{eq30_}
U(z)=\frac{1}{2} \left(V_{\rm ph}(z) -\sqrt{\left(\hbar \Delta +V_{\rm ph}(z) \right)^{2} +4\hbar
^{2} \kappa ^{2} } +\hbar \Omega_{R0} \right).
\end{equation}
The last term in brackets of Eq. \eqref{eq30_} specifies the minimal level of potential energy
$U(z)$, which is equal to zero, $\left. U(z)\right|_{z=0} =0$ at the center of the trap. From
Eq.\eqref{eq30_} under the condition \eqref{eq25_} for photon-like polaritons  one can obtain a
simple expression for power-law potential
\begin{equation} \label{eq31_}
U(z)\simeq U_{\rm pol} \left|z\right|^{\nu },
\end{equation}
where $U_{\rm pol} ={m_{\rm ph} c^{2} \alpha \left(\Omega _{R0} +\left| \Delta \right|
\right)}/{2\Omega _{R0} } $.

Let us examine statistical properties of LB polaritons trapped in the waveguide. From the above
discussion, we can treat the polaritons as one-dimensional ideal bosons confined in the potential
$U(z)$. In the quasiclassical approximation \eqref{eq25_} the density of states approaches
\begin{equation} \label{eq32_}
\rho \left(\varepsilon \right)=\frac{\sqrt{2} }{\pi \hbar } \int _{0}^{l\left(\varepsilon
\right)}\left(\frac{m_{\rm pol}^{} }{\varepsilon -U(z)} \right)^{1/2} dz ,
\end{equation}
where $l\left(\varepsilon \right)=\left(\frac{\varepsilon }{U_{\rm pol} } \right)^{1/\nu} $ is a
characteristic length of localization for LB polaritons with energy $\varepsilon $ -- see
Fig.\ref{fig5}a. The density of states $\rho \left(\varepsilon \right)$ as a function of the
normalized polariton energy $\varepsilon /\hbar \left|\Delta \right|$ is plotted in Fig.\ref{fig5}b
for the approximate (dashed) and exact (solid) potential. Differences occur at polariton energies
of $\varepsilon \sim \hbar \left|\Delta \right|$, i.e. for LB polaritons weakly confined inside the
region $z<z_{1/2} $  one can use power-law approximation \eqref{eq31_} of the trapping potential
$U(z)$.

The total number of LB polaritons $N_{\rm pol} $ is given by
\begin{equation} \label{eq33_}
N_{\rm pol} =N_{0} +\int \frac{\rho \left(\varepsilon \right)d\varepsilon }{\exp
\left[{(\varepsilon -\mu )\mathord{\left/ {\vphantom {(\varepsilon -\mu ) k_{B} T}} \right.
\kern-\nulldelimiterspace} k_{B} T} \right]-1}  ,
\end{equation}
where $N_{0} $ is the number of ground state polaritons, $\mu $ is chemical potential.

We find the critical temperature $T_{C}$, for which the ground state occupation becomes
macroscopically by solving Eq.\eqref{eq33_} at $\mu =0$.  For the onset of Bose-Einstein
condensation one finds (cf. \cite{3})
\begin{equation} \label{eq34_}
k_{B} T_{C} =\left[\frac{\pi \hbar N_{\rm pol} \nu U_{\rm pol}^{1/\nu } }{\sqrt{2m_{\rm pol} }
F(\nu )\Gamma (x)\zeta (x)} \right]^{2\nu /(2+\nu )} ,
\end{equation}
where $F(\nu )=\int _{0}^{1}\frac{t^{1/\nu -1} dt}{\sqrt{1-t} } $; $\Gamma (x)$ and $\zeta (x)$ are
the gamma and the Riemann (zeta) functions taken at $x=1/\nu +1/2$, respectively. Below the
critical temperature the occupation of the ground state is then determined by
\begin{equation} \label{eq35_}
N_{0} =N_{\rm pol} \left[1-\left( \frac{T}{T_{C} }\right)^{1/\nu +1/2}  \right].
\end{equation}

\begin{figure}
\includegraphics[scale=0.45]{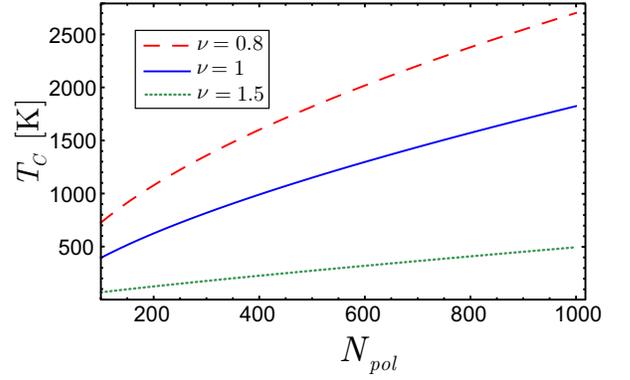}
\caption{\label{fig6}(Color online). Critical temperature \textbf{$T_{C} $} versus  number of
polaritons $N_{\rm pol}$.}
\end{figure}

In the experiment, the average number  of LB polaritons $N_{\rm pol} =\sum _{\vec{k}} \left\langle
\hat{\Xi }_{\vec{k}}^{\dag } \hat{\Xi }_{\vec{k}} \right\rangle \approx N_{\rm ph} $ can be
estimated using the photon-like  character of polaritons  in the perturbation limit.  Notice that a
polaritonic model is valid under the low excitation density limit for which the average number of
photons $N_{\rm ph} $ is essentially smaller than the average number of  atoms $N_{\rm at} $. Using
experimentally accessible rubidium atom densities, that are $n_{\rm at} =N_{\rm at}/V_{M} =10^{16}
{\rm cm}^{-3} $, cf. \cite{19}, and taking the occupation volume $V_{M} $ of the lowest photonic
mode (TM$_{01}$ mode) to be $V_{M} \approx 0.5\, \mu m^{3} $  we find an average number of $N_{\rm
at} =5000$ atoms in the BWC, thus limiting the experiment to a few hundred polaritons.
In Fig.\ref{fig6} the critical temperature $T_{C} $ as a function of LB polariton number
$N_{\rm pol} $ is shown. High temperature BEC, as it follows from Eq.\eqref{eq34_} and
Fig.\ref{fig6} can be achieved for an experimentally feasible number of polaritons. Since the
function $\zeta (x)$ is a diverging at   $x=2$ (see Eq.\eqref{eq34_} and \cite{23_}), the critical
temperature $T_{C} $  vanishes for increasing trapping power parameter $\nu $ - see Fig.\ref{fig6}.
From Eq.\eqref{eq34_} we find a critical temperature which drastically increases with increasing
$\alpha$-parameter. However, the values of the parameter are limited by inequality \eqref{eq2_} for
our problem.

Notably, the thermal de Broglie wavelength $\Lambda_{T} = \left[ \frac{2\pi \hbar ^2}{m_{\rm pol}
k_{B} T} \right]^{1/2} $ at the experimentally explored temperatures of the atomic gas $T=530$ K is
macroscopically large, i.e. $\Lambda = 1.89$ $\mu m$    and comparable with the magnitude of
characteristic length $d_{1}$  of photonic field localization.

\section{POLARITON BEC PROPERTIES AT $\nu =1$}

Let us examine  LB polariton condensate properties at sufficiently ``low'' temperatures such as $T
\ll T_{C} $ which is relevant to the linear trapping potential $U(z)$ obtained at  $\nu =1$. In
this case the parameter $U_{\rm pol} = U\left( z\right) / \left| z \right|$ may be physically
interpreted  as a force acting on polaritons in the waveguide cavity. Macroscopic LB polariton BEC
properties  can be found with the help of a quantum field theory approach. In particular, the
Lagrangian density $L$ for the system described by Hamiltonian \eqref{eq29_} looks like
\begin{equation} \label{eq36_}
L=\frac{i\hbar }{2} \left(\psi \frac{\partial \psi ^{*} }{\partial t} -\psi ^{*} \frac{\partial
\psi }{\partial t} \right)+\frac{\hbar ^{2} }{2m_{\rm pol}} \left|\frac{\partial \psi }{\partial z}
\right|^{2} +U_{\rm pol} \left|z\right|\left|\psi \right|^{2} ,
\end{equation}
where  $\psi (z,t)\equiv \left\langle \hat{\Xi }(z,t)\right\rangle $ is classical polariton
condensate ground state wave function normalized as $\int _{-\infty }^{+\infty }\left|\psi
(z,t)\right|^{2} dz =1$.   Lagrangian  density $L$ implies non-stationary Schr\"odinger equation
for polaritons in \textit{coordinate} representation in the form
\begin{equation} \label{eq37_}
i\hbar \frac{\partial \psi (z,t)}{\partial t} =\left[-\frac{\hbar ^{2} }{2m_{\rm pol} } \frac{d^{2}
}{dz^{2} } +U_{\rm pol} \left|z\right|\right]\psi (z,t).
\end{equation}
To obtain a stationary solution we make a substitution $\psi (z,t)=e^{-i\mu t/\hbar } \psi (z)$,
where $\mu $ is a chemical potential. Then we have
\begin{equation} \label{eq38_}
\frac{\hbar ^{2} }{2m_{\rm pol} } \frac{d^{2} \psi (z)}{dz^{2} } +\left(\mu -U_{\rm pol}
\left|z\right|\right)\psi (z)=0.
\end{equation}

The LB polariton ground state wave function is expressed  through the Airy function as, cf.
\cite{28}:
\begin{equation} \label{eq39_}
\psi (z)=\frac{1.308}{\sqrt{d_{\rm 1, pol} } } {\rm Ai}\left(\frac{\left|z\right|}{d_{\rm 1, pol} }
+a'_{1} \right),
\end{equation}
where $d_{\rm 1, pol} = \left[ \frac{\hbar^{2}}{2 m_{\rm pol} U_{\rm pol}}\right] ^{1/3} $
specifies a characteristic  scale of spatial (longitudinal) localization of the polariton
condensate for a linear trapping potential; $a^{ \prime} _{1} \approx -1.0188$ is the first zero of
the Airy function derivative. It is interesting to note that the characteristic length $d_{\rm 1,
pol}=d_1$ is the same for the photons and photon-like polaritons trapped in the BWC
due to relation $m_{\rm pol}U_{\rm pol}=m_{\rm ph}U_{\rm ph}$, that is true under the condition \eqref{eq25_}, see also Eqs. (\ref{eq16}),
(\ref{eq31_}). Since LB polaritons are completely photon-like it is possible to conclude that the
photon mode $n=0$ reflects the behaviour of polaritonic probability density
$\left| \psi _{n} \right|^{2} \equiv \left| \psi \right|^{2}$. The chemical potential $\mu $ can be easily found from
Eqs.\eqref{eq38_}-\eqref{eq39_} and given by
\begin{equation} \label{eq40_}
\mu =1.0188\, U_{\rm pol} d_{1} .
\end{equation}

To investigate the reaction of the BEC to a small disturbance induced from the outside we
study the dynamics of the BEC when the initial wave function has been compressed or
stretched compared to the equilibrium wave function. In this case one expects slow oscillations of the
BEC around the equilibrium state. The dynamics of the polariton BEC induced by a disturbance can
be found by means of the variational approach for solving Eq.\eqref{eq37_}. In particular, we take
the Airy trial function for the ground state (cf. \cite{29}):
\begin{equation} \label{eq42_}
\psi (z,t)=\frac{{\mathbb N}}{\sqrt{D(t)} } {\rm Ai}\left(\frac{\left|z\right|}{D(t)} +a'_{1}
\right)e^{ib(t)\left|z\right|}   ,
\end{equation}
${\mathbb N}$ is a normalization constant. In Eq.\eqref{eq42_} the time-dependent function $D(t)$
specifies the width of the wave function, and $b(t)$ characterizes a related wave function
curvature. Inserting Eq.\eqref{eq42_} into Eq.\eqref{eq36_}  it is possible to obtain  an effective
Lagrangian $\bar{L}=\int \limits_{- \infty}^{+ \infty} L dz$ by averaging Lagrangian density $L$ as
\begin{equation} \label{eq43_}
\bar{L}=-\frac{2\hbar a'_{1} }{3} \frac{db}{dt} D-\frac{\hbar ^{2} a'_{1} }{6m_{\rm pol} }
\frac{1}{D^{2} } +\frac{\hbar ^{2} }{2m_{\rm pol} } b^{2} -\frac{2U_{\rm pol} a'_{1} }{3} D.
\end{equation}

The effective Lagrangian $\bar{L}$ leads to a Newton-like equation for condensate width function
\begin{equation} \label{eq44_}
\frac{d^{2} D}{dt^{2} } =\frac{3\hbar ^{2} }{4\left|a'_{1} \right|m_{\rm pol}^{2} D^{3} }
-\frac{3U_{\rm pol} }{2\left|a'_{1} \right|m_{\rm pol} }.
\end{equation}
For further processing it is useful to introduce a new dimensionless variable for the wave function
width $w=D/d_{1} $ and the rescaled time  $\tau =\omega _{\rm pol} t$, where $\omega _{\rm pol} $
is a characteristic frequency of polariton trapping defined as (cf. \eqref{eq16_})
\begin{equation} \label{eq45_}
\omega _{\rm pol} =\frac{\sqrt{3} \hbar }{2\sqrt{\left|a'_{1} \right|} m_{\rm pol} d_{1}^{2} }
\simeq \left(\frac{3^{3/2} U_{\rm pol}^{2} }{2\left|a'_{1} \right|^{3/2} \hbar m_{\rm pol} }
\right)^{1/3}.
\end{equation}

\begin{figure}
\includegraphics[scale=0.5]{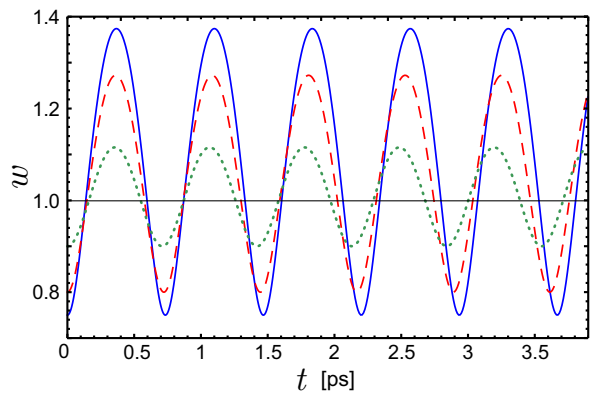}
\caption{\label{fig7}(Color online) Normalized wave function width $w$ versus time $t$. The
parameters are the following: $\nu =1$, $R_{0} \approx 0.3$ $\mu {\rm m}$ and $\alpha^{1/\nu}
=0.0005$ $\mu {\rm m}^{-1} $. An initial condition is  $w\left(\tau =0\right)=0.75$ for solid
(blue) curve, $w\left(\tau =0\right)=0.8$ for  dashed (red) curve,  and $w\left(\tau =0\right)=0.9$
for dotted (green) curve. In all cases $\dot{w}\left(\tau =0\right)=0$. The horizontal line  $w=1$
corresponds to the stationary solution of Eq.\eqref{eq46_}.}
\end{figure}

The dimensionless equation for the new variables takes the form
\begin{equation} \label{eq46_}
\frac{\partial ^{2} w}{\partial \tau ^{2} } =\frac{1}{w^{3} } -1 .
\end{equation}
The solution of Eq.\eqref{eq46_} in the absence of polariton trapping is $w=\sqrt{1+\tau ^{2} } $;
describing the spreading of a free polariton wave packet with Airy shape -- cf. \cite{28}. In the
presence of polariton trapping the analytical  solution of Eq.\eqref{eq46_} is much more
complicated. Figure \ref{fig7} demonstrates the temporal dynamics of the normalized polariton
condensate wave function width $w$, using the Rubidium parameters described in section
\ref{section_numerics} (cf. \cite{19,20,21}), which shows oscillations around the equilibrium
solution. The characteristic frequency estimated for experimental conditions  is $\omega _{\rm pol}
/2\pi =0.817$ THz. It is important that the value of $\omega _{\rm pol} $ should satisfy the
quasiclassical condition \eqref{eq25_} discussed above. Completely neglecting the decay rate, the
polaritonic system exhibits periodical behavior similar to an atomic BEC in a harmonic trap -- cf.
\cite{29}. The frequency $\Omega_{\rm osc}$ for small amplitude oscillations (see Fig.\ref{fig7})
can be obtained by linearizing Eq.\eqref{eq46_} around the stationary state $w=1$ is found to be
$\Omega _{\rm osc} =\sqrt{3} \omega _{\rm pol} $. The polariton wave function is stable everywhere
in this case.

\section{CONCLUSIONS}

Let us briefly summarize the results obtained. We have considered the problem the thermodynamics
for polaritons emerging due to the interaction of two-level rubidium atoms with an optical field in
the presence of OCs with buffer gas particles. We assume the optical field to be
trapped inside a biconical waveguide filled with Rubidium atoms, and we find conditions for
effective trapping of the coupled atom-light system. In particular, we find a phase transition to a
BEC of polaritons if relation \eqref{eq25_} is fulfilled, a condition that substantially depends on
a cavity mode structure and $\alpha$-parameter characterizing slow changing of waveguide radius
along longitudinal coordinate. We analyze the problem of BEC formation for lower branch atomic
polaritons trapped in the BWC under the quasiclassical approach (\ref{eq17}),
(\ref{eq25_}), and we find that atomic polaritons formed in the waveguide may be treated
as 1D ideal gas of bosonic quasiparticles in the framework of current experimental
results achieved for the thermalization of coupled atom-light states in the presence of  strong atom-field coupling  regime.

Even though the character of the dressed-state polaritons discussed in the paper is nearly
photon-like, they represent mixed states of photons and excitations of two-level atomic system. In
this sense they are physically close to exciton-polaritons obtained in semiconductor microcavities,
cf. \cite{13,14}. However, our estimations show that the critical temperature of the BEC phase
transition can exceed the temperature of 530 K currently used in experiments, thus being high
enough to be observed using appropriate waveguide parameters and cavity modes. Such high critical
temperatures can not be achieved for exciton-polaritons in the  current narrow-band semiconductor
microstructures due  to exciton ionization effects. The main reason for the high transition temperature is the photon-like character
of the polaritons, i.e. their low effective mass. This, together with the thermalization due to
optical collisions, may allow to experimentally observe a high temperature phase transition of
dressed state polaritons using realistic parameters.

In a more detailed study we have investigated the case of a biconical cavity leading to a linear
trapping potential of polariton gas. We are aware that the transverse decoherence rate needs to be
studied in more detail for the considered system to ensure the strong coupling limit. Using
variational techniques we find the ground state wavefunction to be Airy-shaped, where the width $w$
exhibits small amplitude oscillations in time around a stationary state when the BEC is excited,
with a characteristic period in the picosecond domain for current experiments. The results obtained
indicate that the observation of a corresponding thermodynamic phase transition is possible already
in current experimental setups, given the described waveguide structures are prepared with the
required accuracy.

\section*{ACKNOWLEDGMENTS}
This work was supported by RFBR Grants No. 10-02-13300, No. 11-02-97513 and No. 12-02-90419 and by
Russian Ministry of Education and Science under Contracts No. 16.518.11.7030, No. 2.4053.2011 and
No. 3008.2012.2, and by the Deutsche Forschungsgemeinschaft (DFG) under contract No.
436RUS113/995/0-1.

\end{document}